# Crosstalk in Inductive Coupling Communications for Multi-Stacked Chips

Abdullah Saeed Alghotmi
Computer Science Department
Faculty of Computing and
Information
Al-Baha University, Saudi Arabia
aalghotmi@bu.edu.sa

*Abstract—* This paper simulates and analyses the crosstalk interference in off-chip and on-chip communications in multi-stacked chips where Inductive Coupling Links (ICLs) are the communication medium. Crosstalk impacts data signals, leading to errors in the receiver when adjacent transmitters communicate simultaneously. We explored techniques to reduce crosstalk, including varying the coil geometry, changing the coil arrays, and using time-interleaving multiplexing. We propose a 1-of-4 technique, where two bits of data are coded into four bits, always including a '1'. Using an electromagnetic simulator, we described crosstalk behaviour and applied the results to Matlab Simulink to obtain bit-error rates (BER). Simulation results showed that the interference-to-signal ratio (ISR) decreased gradually with increased horizontal coil separation. Additionally, we observed that changing the layout arrays for the coils plays a vital role in reducing crosstalk. The 4-phase time interleaving multiplexing and 1-of-4 technique significantly reduced crosstalk at the expense of bandwidth.

## I. Introduction and Background

Shrinking the size of chips and doubling the transistors' capacity every two years, in line with Moore's law, has led to research into multi-stacked chips [1]. These chips are vertically stacked on top of each other, forming what is known as 3D-integrated circuits (3D-ICs). The 3D-ICs are increasingly used in high-performance computing, IoT devices and medical equipment due to their compact size and enhanced performance. Linking those stacked chips can be wired-based or wireless. The state-of-the-art for the wired approach is Through-silicon-Via (TSV), but its manufacturing varies through several costly stages, making it cost more and prone to defects [2]. The other approach is wireless where chips communicate using capacitive or inductive coupling links. The capacitive coupling link has some limitations due to technology integration while inductive coupling can communicate in a wide area and is not limited to a specific integration style [3]. 3D-ICs using Inductive Coupling Links (ICLs) utilise magnetic fields to transfer data with a series of current pulses. An input pulse on the transmitter (primary coil) is received in the receiver (secondary coil) as two complementary pulses, corresponding to the build-up and decay of the primary current, respectively. The operation of ICLs is fundamentally based on Faraday's Law of Induction, which states that a changing magnetic field within a closed loop induces an electromotive force (EMF) in the wire. This principle is further described by Maxwell's equations, which provide a comprehensive framework for understanding electromagnetic fields and their interactions.

There exist different signalling schemes in ICL communication. For example, bi-phase signalling maps the binary data to a series of current pulses where "1's" are mapped to positive and "0's" to a negative polarity. Another scheme used in ICL is non-return to zero (NRZ) signalling. NRZ is the most applied signalling scheme in inductive coupling in 3D-ICs. It operates similarly to bi-phase; however, in NRZ, rising and falling edges are mapped to pulses with positive and negative polarity. Both coding schemes are shown in Figure 1.

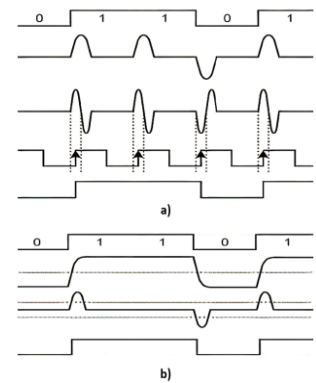

*Fig 1. Pulse representations in ICL communications using (a) Bi-phase, (b) Non-return-to-zero NRZ [4]*

The benefits of utilising ICL in 3D-ICs are that it can communicate more widely than other vertical communication approaches and is cheaper in manufacturing, making it more favoured [5]. However, ICL suffers from communication interference between the adjacent coils, which is a crucial concern causing timing or noise disruption. This interference results in jitter in the time domain, voltage noise in the frequency domain, or both [6]. This phenomenon is called crosstalk, caused by mutual inductance between adjacent coils. Numerous techniques have been established to reduce crosstalk, including placing the coils further apart, which has a significant role in manufacturing [7], selecting coding or a signalling scheme that suits the characteristics of inductive coupling communications [4], and using time interleaving multiplexing [8].

This work utilised the time-interleaving multiplexing technique and 1-of-4 coding scheme. Both techniques were used in different measurements and setups. Different scenarios, including on-chip and off-chip crosstalk, were examined to determine the unwanted coupling the receivers suffer. In the experiment, we designed the coil structures and measured the crosstalk behaviour of the coils with the Ansys High-frequency Structure Simulation (HFSS) simulator to extract several parameters for the Simulink simulator. We swept different setups for different parameters to examine the ISR in two and four pairs of coils. Then, we abstracted the communication blocks in Simulink to calculate the bit-error rate (BER) without any deduction technique. Lastly, we compared the BER results from the original signal to the time interleaving multiplexing technique and 1-of-4 coding.

## II. RELATED WORK

Crosstalk is a significant issue in ICLs for 3D-ICs, affecting signal integrity and increasing error rates. Recent research has focused on various aspects of crosstalk, including theoretical models, simulation methods, and mitigation strategies.

### A. Theoretical Models and Simulation Methods

Several studies have developed theoretical models to understand and predict crosstalk in ICLs. For instance, [8] presented a comprehensive analysis of crosstalk in high-density multi-hop inductive coupling interfaces for 3D-stacked memory. They used electromagnetic simulations to measure the interference-to-signal ratio (ISR) and bit-error-rate (BER) under various configurations.

The work in [5] proposed a time-interleaving technique to reduce crosstalk in 3D-ICs. Their work involved narrowing the communication distance between coils vertically and analysing the resulting ISR and jitter. Their findings highlighted the trade-off between noise tolerance and bandwidth, demonstrating that increased noise tolerance leads to reduced bandwidth.

### B. Crosstalk Mitigation Techniques

The Recent advancements in crosstalk mitigation have focused on improving the design of coil layouts and employing sophisticated signalling schemes. The authors in [6] explored the use of different signalling schemes, such as single-phase modulation (SPM), bi-phase modulation (BPM), and non-return to zero (NRZ) signalling, to reduce power consumption in ICLs. They also introduced a new phase code modulation technique that effectively reduces power consumption but has a lower resolution, leading to timing uncertainties.

The effects of crosstalk on power distribution networks in inductive and capacitive coupling in 3D-ICs were investigated in [3]. The authors used both analytical and numerical methods to model the mutual inductance between coils and proposed design modifications to mitigate crosstalk. Their study provided insights into the impact of coil separation and layout on crosstalk levels.

### C. Innovations in Coil Design

Innovative coil designs have also been explored to minimise crosstalk. The work in [4] developed a simultaneously bidirectional inductively coupled link in a 0.13-μm CMOS technology, demonstrating significant reductions in crosstalk through improved coil geometries. Their experimental results showed that specific geometric adjustments could lead to lower ISR and BER.

Additionally, recent work [9] focused on optimising the transmitter and receiver coils layout to minimise crosstalk. They analysed various configurations and determined that placing transmitters and receivers on separate dies significantly reduces crosstalk compared to placing them together on the same die.

The study in [10] conducted a comprehensive analysis of crosstalk in high-bandwidth 3D-stacked memory using a multi-hop inductive coupling interface. They identified two main sources of crosstalk: concentric coils and adjacent coils. To address these issues, they proposed two countermeasures: shorted coils and 8-shaped coils. These techniques demonstrated significant improvements in area efficiency and crosstalk reduction through detailed simulations.

### D. Contribution

ICLs for 3D-ICs have been studied extensively in recent years with an emphasis on understanding and mitigating crosstalk. Theoretical models and simulation methods have provided a deeper understanding of crosstalk mechanisms, while advancements in signalling schemes and coil designs have offered practical solutions for reducing interference.

Despite these advancements, there remains a need for comprehensive studies that combine various mitigation techniques and provide a holistic approach to crosstalk reduction. This paper contributes to the following:

• **Simulating and Analysing Crosstalk in Various Setups**: We identify the worst-case scenarios through detailed simulations using HFSS and validate the results using Matlab Simulink.

• **Introducing and Evaluating Mitigation Techniques**: We evaluate the effectiveness of time the introduced encoding scheme 1-of-4 alongside existing techniques like time interleaving in reducing ISR and BER, providing a comparative analysis not extensively covered in prior research.

• **Proposing Practical Solutions**: By integrating multiple mitigation strategies, our work offers practical guidelines for designing ICLs in 3D-ICs to achieve optimal performance with minimal crosstalk by analysing the effectiveness of each technique through detailed simulations.

## III. SYSTEM MODEL

Crosstalk in inductive coupling communications within 3D-ICs is a critical issue that can degrade signal integrity and increase error rates. This section presents a theoretical model to quantify crosstalk effects in ICLs.

### A. Theoretical Model

*1) Faraday's Law and Maxwell's Equations:* The behaviour of inductive coupling can be explained using Faraday's Law, which states that the induced voltage in a coil is proportional to the rate of change of the magnetic flux through the coil. Maxwell's equations provide a broader framework for understanding the interactions of electric and magnetic fields. The relevant equations are:

$$\vec{\nabla} \times \vec{E} = -\frac{\partial \vec{B}}{\partial t} \quad (Faraday's\ Law) \quad (1)$$

$$\vec{\nabla} \times \vec{H} = \vec{J} + \frac{\partial \vec{D}}{\partial t} \quad (Ampère's Law\ with\ Maxwell) \quad (2)$$

*2) Crosstalk Model:*

*Mutual Inductance:* Mutual inductance ($M$) between two coils is defined as:

$$M = k\sqrt{L_1 L_2} \quad (3)$$

where $k$ is the coupling coefficient, and $L1$ and $L2$ are the inductances of the primary and secondary coils, respectively.

*Induced Voltage*: The voltage induced in the secondary coil ($V2$) due to a changing current in the primary coil ($I1$) is given by:

$$V_2 = M \frac{dI_1}{dt} \quad (4)$$

*Crosstalk Interference-to-Signal Ratio (ISR)*: ISR is a key parameter to measure the crosstalk effect, defined as the ratio of the interference voltage to the signal voltage:

$$\text{ISR} = \frac{V_{\text{crosstalk}}}{V_{\text{signal}}} \quad (5)$$

*Impact of Coil Separation and Layout*: The impact of coil separation ($d$) and layout on mutual inductance can be expressed as:

$$M \propto \frac{1}{d^n} \quad (6)$$

where $n$ depends on the specific geometry and spacing of the coils.

*Bit-Error Rate (BER)*: Crosstalk can also be quantified in terms of the bit-error rate (BER), which is affected by ISR. For a given communication system, BER can be estimated using:

$$\text{BER} \approx Q\left(\sqrt{\frac{2E_b}{N_0(1+\text{ISR})}}\right) \quad (7)$$

*B. Simulation-Based Quantification of Crosstalk*

To quantify crosstalk, we used both analytical and numerical methods. Analytical methods involve solving the Neumann formula for inductance, which can be complex for non-trivial geometries:

$$M = \int_{\text{coil 1}} \int_{\text{coil 2}} \frac{\mu_0}{4\pi} \frac{d\mathbf{l_1} \cdot d\mathbf{l_2}}{|\mathbf{r_1} - \mathbf{r_2}|} \quad (8)$$

where $\mu_0$ is the permeability of free space, $dl_1$ and $dl_2$ are infinitesimal elements of the coils, and $r_1$ and $r_2$ are position vectors of these elements.

Numerical methods, such as those implemented in electromagnetic simulation software like Ansys HFSS, can model the precise geometries and materials of the coils to calculate mutual inductance accurately. Ansys HFSS allows for detailed electromagnetic simulations that consider the physical layout, materials, and operating conditions of the inductive coupling system. By performing these simulations, we can assess the crosstalk behaviour under various configurations and validate our theoretical models.

*TABLE I. METRICS USED IN COIL SEPERATION*

| Metrics | Ref[4] | Ref[11] | This work |
|---|---|---|---|
| Horizontal distance | 20um | 40 | 10 & 60um |
| Vertical distance | 300um | 88.1um | 106um |
| CMOS technology | 0.35mm | 0.35mm/65 & 28nm | 65nm |
| Inductor diameter | 100x100um | 250x250um | 250x250um |
| Adhesive | 10um | 20um | 20um |
| Number of turns | 3 | 5 | 5 |
| Array layout | 5 x 5 | 1 x 3 | 2 x 2 |
| Trace Spacing | 0.5um | 1um | 1um |
| Trace Width | 1um | 9um | 1, 3, 5um |

IV. CROSSTALK SIMULATIONS AND RESULTS

*A. Simulation Setup*

The coil structures are designed within the simulator, with excitations made based on a lumped source. This lumped port power source is generic for computing Scatter Parameters (s-parameters). S-parameters are critical for modelling the coupling between high-speed frequency signals and analysing the coupling and voltage transfer of the coils. The HFSS simulator simulates the coils' behaviour, following the simplified model shown in Figure 2. Crosstalk is measured and analysed for both on-chip and off-chip scenarios by simulating the magnetic field interactions between transmitters (TXs) and receivers (RXs) coils in two stacked chips, as illustrated in Figure 3.

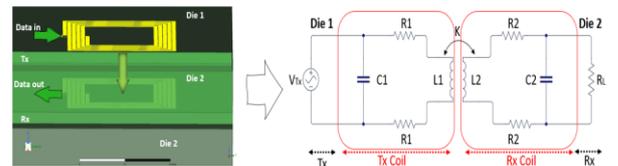

*Fig 2. Simplified circuit model for an ICL*

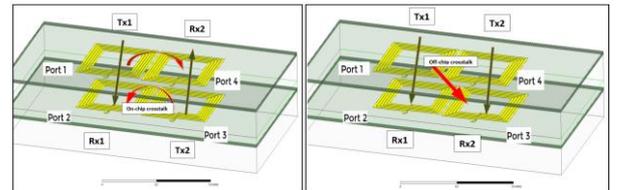

*Fig 3. Crosstalk in on-chip and off-chip scenarios*

HFSS utilises voltage levels to compute the s-parameters, characterising the radio-frequency components and calculating properties such as gain, loss, and other linear network parameters. The s-parameter data for each combination of coils is exported and fed into Matlab Simulink using block models as shown in Figure 4.

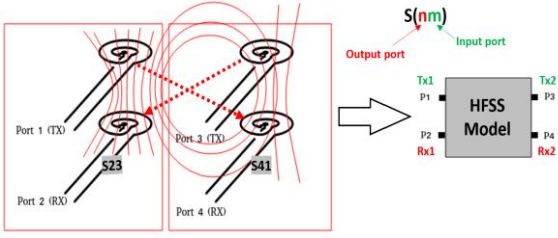

Fig 4. Crosstalk in on-chip and off-chip scenarios

B. Simulation Process

The simulation involves setting up current sources to simulate signal inputs to the primary coils and running the simulations to compute electromagnetic field distributions, induced voltages, and currents within the coils. These simulations are conducted using Ansys HFSS to determine the S-parameters, which are then imported into Matlab Simulink for further analysis as in Figure 5.

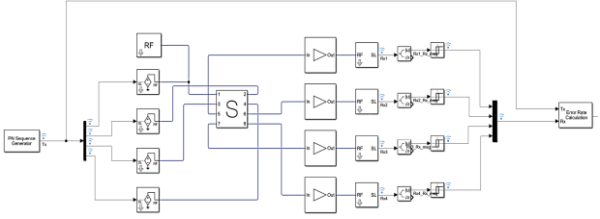

Fig 5. S-parameters imported in Simulink model

C. Simulation and Analysis

Guided by the theoretical model, we simulate different scenarios to analyse crosstalk using Ansys HFSS and Matlab Simulink.

- **Single Pair of Coils**: The coupling for different distances and its impact on ISR is presented in Table I.
- **Multiple Pairs of Coils**: ISR and BER for configurations with two and four pairs of coils are shown in Table II.
- **Crosstalk Mitigation Techniques**: The effectiveness of time interleaving and 1-of-4 encoding schemes is validated through simulation results, demonstrating significant improvements in ISR and BER.

ISR is an essential parameter for measuring the crosstalk interference in HFSS. Previous works have measured ISR and BER [7], while other studies have measured jitter and displayed eye diagrams [12]. This paper measures ISR in HFSS and BER in Simulink based on several coil metrics and setups as discussed in the following subsections. We focus on ISR and BER as they provide precise, quantitative metrics directly related to crosstalk interference and signal integrity, suitable for our simulation tools. Additionally, we introduced the 1-of-4 encoding scheme which showed significant improvements in BER.

*1) On-Chip Coupling Crosstalk:* The on-chip coupling crosstalk signal was measured as $S_{21} = 0.12$. The relationship between $S_{21}$ and the received voltage ($V_2$) is given by:

$$S_{21} = \frac{V_2}{V_1} \quad (9)$$

Given $V_1 = 1.2\,\text{V}$,

$$0.12 = \frac{V_2}{1.2\,\text{V}} \quad (10)$$

Solving for $V_2$,

$$V_2 = 1.2\,\text{V} \times 0.12 = 0.144\,\text{V} = 144\,\text{mV} \quad (11)$$

Thus, the received signal for on-chip coupling is 144 mV.

*2) Off-Chip Coupling Crosstalk:* For off-chip coupling crosstalk, the signal was measured as $S_{42} = 0.009$. The relationship between $S_{42}$ and the received voltage ($V_4$) is given by:

$$S_{42} = \frac{V_4}{V_2} \quad (12)$$

Given $V_2 = 1.2\,\text{V}$,

$$0.009 = \frac{V_4}{1.2\,\text{V}} \quad (13)$$

Solving for $V_4$,

$$V_4 = 1.2\,\text{V} \times 0.009 = 0.0108\,\text{V} = 10.8\,\text{mV} \quad (14)$$

Thus, the received crosstalk for off-chip coupling is 10.8 mV.

The received voltage for on-chip coupling is 144 mV, of which 10.8 mV is crosstalk. Since all coils are equal in size, these results are consistent with other parameters. Additionally, we converted the values to dB for convenience in communication interference measurements, as shown in Table II.

D. Results Analysis

The results showed that varying the number of coil pairs and their configurations significantly impacts crosstalk behaviour. Increasing the number of coil pairs increased the complexity of the mutual inductance interactions, leading to higher ISR values. Both time interleaving and the novel 1-of-4 encoding showed improvements in BER, demonstrating that choosing the appropriate encoding scheme can significantly impact crosstalk reduction and overall system performance.

Table II. RELATIONSHIP BETWEEN ISR AND HORIZONTAL SEPARATION

| Horizontal Spacing | On-chip Crosstalk (dB) | Off-chip Crosstalk (dB) |
|---|---|---|
| 10 µm | -26.30 | -41.62 |
| 20 µm | -27.92 | -42.16 |
| 90 µm | -38.64 | -43.61 |

V. DISCUSSION

Coupling depends on the mutual inductance, and the mutual inductance depends on the inductance of each coil and the trace spacing. Additionally, the inductance of each coil depends on the area and the number of turns. As a result, designing the coil is of primary importance. Table I shows the setup parameters considered in designing the coils in this work and other related works. In the following subsections, different setups, including one pair, two pairs, and four pairs of coils, are discussed.

### A. Trace Width

We explored the coupling changes for different width traces, including 6μm, 9μm, and 12μm. The 12μm trace width showed the highest coupling, but practically it hasn't been tested in real applications which leads us to focus on 9μm, the standard research trace width [7]. However, in the future, there is an intention for testing more metrics where the coupling should be higher than 6μm and 9μm. The results from HFSS for the coils parameters and Simulink for the BER are shown in Table III.

*Table III. COUPLING IN DIFFERENT TRACE WIDTHS AND THE BER MEASUREMENTS FROM SIMULINK*

| Trace Width | 0.1GHz | 1GHz | 1.45GHz | 3GHz | BER |
|---|---|---|---|---|---|
| 6μm | -36.0 dB | -20.2 dB | -20.1 dB | -22.9 dB | 30.1% |
| 9μm | -35.6 dB | -19.0 dB | -18.3 dB | -20.3 dB | 20.6% |
| 12μm | -35.4 dB | -18.3 dB | -17.4 dB | -19.0 dB | 11.1% |

### B. Trace Spacing

Simulation results do not change significantly when the spacing between traces is altered without changing the geometry of the coil diameter, as shown in Table IV. We are planning in future work to change the actual geometry of the coil from 250μm x 250μm to greater and change the trace spacing to analyze different results.

*Table IV. COUPLING RATIOS IN DIFFERENT TRACE SPACING AND THE BER MEASUREMENTS FROM SIMULINK*

| Trace Spacing | 0.1GHz | 1GHz | 1.45GHz | 3GHz | BER |
|---|---|---|---|---|---|
| 1μm | -35.6 dB | -19.0 dB | -18.3 dB | -20.3 dB | 20.6% |
| 3μm | -37.1 dB | -19.6 dB | -18.5 dB | -19.7 dB | 12.6% |
| 5μm | -37.8 dB | -19.9 dB | -18.4 dB | -18.9 dB | 9.5% |

### C. One Pair of Coils

We first determined the coupling between trace widths and spacing to calculate the worst-case couple as shown in Figure 6. The coupling for different distances and its impact is presented in Table II.

### D. Two Pairs of Coils

After we determined how much coupling there was between one pair of coils, we placed another pair. We measured and analyzed the coupling in different horizontal setups. As shown in Figure 7, the closer the coils are, the greater the coupling and potential crosstalk between them. The effects of the two pairs in on-chip and off-chip scenarios are shown in Figure 6, where the x-axis is the horizontal separation, and the y-axis is the crosstalk ratio.

We considered the worst-case scenario for crosstalk coils separation, which is 10μm = 260μm middle to middle. As a result of the experiment, it is clear that the arrangement of the coils significantly impacts the crosstalk.

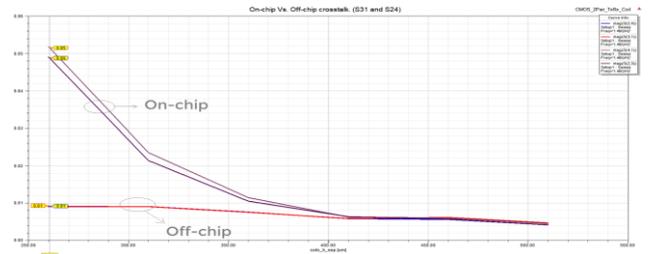

*Fig 6. On-chip Vs. Off-chip Crosstalk*

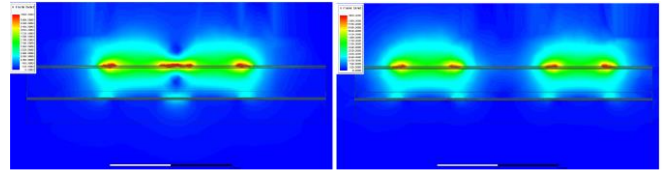

*Fig 7. EM behaviour for two adjacent pairs in two different spacing*

### E. Four Pairs of Coils

Four pairs of coils are laid in a 2x2 pattern where we have two scenarios as in the previous section, on-chip crosstalk and off-chip as shown in Figure 8. We took the measurements from HFSS to show how much crosstalk is within different scenarios. Then, we extracted the data from HFSS to measure the BER in Simulink where each coil transmits 1 bit of data simultaneously. After we calculated the BER, we applied the crosstalk deduction techniques.

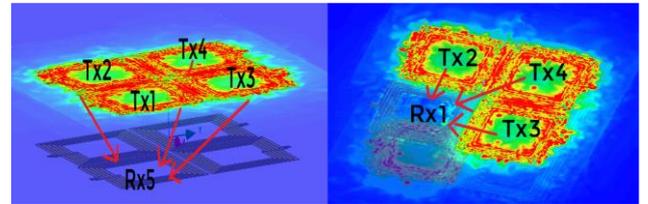

*Fig 8. 4-phase time interleaving model in Simulink (a) off-chip, (b) on-chip*

### F. Crosstalk Mitigation Techniques

4-Phase Time Division Multiplexing Technique and 1-of-4 Encoding Scheme: 4-phase time interleaving multiplexing reduces the BER at the expense of the bandwidth where it allows one coil to transmit 1 bit at a time. The BER was reduced from 21% to about 1%, as shown in Table V. Consequently, the bandwidth dropped to 10Gb/s, whereas without using this technique, it was 40Gb/s. We also examined a different technique, which is 1-of-4 coding. Simulation shows bandwidth improvement while maintaining the crosstalk at a reduced level, as shown in Table V. The introduced encoding scheme 1-of-4 is a compromise solution to reduce crosstalk by avoiding many coils transmitting at the same time, but not losing so much bandwidth as the 4-phase.

*Table V. SIMULINK RESULTS FOR BASIC (ACTUAL OUTPUT), 4-PHASE, AND 1-OF-4 TECHNIQUES*

| Techniques | Bits | bits/s | BER |
|---|---|---|---|
| Basic | 4 | 40 Gb/s | ~21% |
| 4-Phase | 1 | 10 Gb/s | ~1% |
| 1-of-4 | 2 | 20 Gb/s | ~1% |

*G. Implications and Future Work*

Our findings highlight the importance of coil design and spacing in mitigating crosstalk in 3D-ICs. The simulation results demonstrate that adjusting the trace width and spacing can significantly impact crosstalk levels and BER. Additionally, the choice of coding scheme plays a critical role in managing crosstalk. Our work with the 1-of-4 coding scheme has shown promising results in reducing crosstalk while maintaining bandwidth efficiency.

Future work will focus on further optimising these parameters and exploring additional crosstalk reduction techniques. Specifically, we plan to investigate alternative coding schemes that could offer better performance under different operating conditions. Moreover, further detailed simulations will be crucial to validate the proposed solutions and ensure the feasibility in another circuit simulator (ex, Cadence).

## VI. Conclusion

Mutual inductance leads to crosstalk between adjacent ICLs, and this work measured the crosstalk for on-chip and off-chip scenarios using two simulators, Ansys HFSS and Matlab Simulink, presenting results in terms of magnetic behaviour, ISR, and BER representation. We measured the voltage between a pair of coils and extended the analysis to configurations with two and four pairs of coils. The time interleaving technique demonstrated significant crosstalk reduction across various scenarios. Notably, the introduction of the 1-of-4 encoding scheme showed promising results in further reducing crosstalk and improving BER.